\documentclass[twocolappendix,appendixfloats,]{emulateapj}

\usepackage[colorlinks = true, linkcolor = blue, urlcolor  = blue, citecolor = blue, anchorcolor = blue]{hyperref}
\usepackage{hyperref}
\usepackage{apjfonts}
\usepackage{rotating}
\usepackage{amsmath}
\usepackage{color}
\usepackage{graphicx}
\usepackage[normalem]{ulem}
\usepackage[dvipsnames]{xcolor}
\usepackage{CJK}

\newcommand{\thetae}{\theta_{\rm E}}

\newcommand{\pie}{\pi_{\rm E}}

\definecolor{brown}{rgb}{0.59, 0.29, 0.0}
\definecolor{darkgreen}{rgb}{0.0, 0.42, 0.24}
\definecolor{darkblue}{rgb}{0.01, 0.31, 0.59}
\definecolor{darkblue}{rgb}{0.0, 0.25, 0.42}

\shorttitle{One Planet or Two Planets? }
\shortauthors{Han et al.}

\begin{document}

\title{
One Planet or Two Planets? The Ultra-sensitive Extreme-magnification
Microlensing Event KMT-2019-BLG-1953}

\author{
Cheongho~Han$^{001,101}$, 
Doeon~Kim$^{001}$,
Youn~Kil~Jung$^{002}$, 
Andrew~Gould$^{003,004,101}$,
Ian~A.~Bond$^{005,102}$,
\\
(Leading authors),\\
Michael~D.~Albrow$^{006}$, 
Sun-Ju~Chung$^{002,007}$,  
Kyu-Ha~Hwang$^{002}$, 
Chung-Uk~Lee$^{002}$, 
Yoon-Hyun~Ryu$^{002}$, 
In-Gu~Shin$^{002}$, 
Yossi~Shvartzvald$^{008}$, 
Jennifer~C.~Yee$^{009}$, 
Weicheng~Zang$^{010}$,
Sang-Mok~Cha$^{002,011}$, 
Dong-Jin~Kim$^{002}$, 
Hyoun-Woo~Kim$^{002}$, 
Seung-Lee~Kim$^{002,007}$, 
Dong-Joo~Lee$^{002}$, 
Yongseok~Lee$^{002,011}$, 
Byeong-Gon~Park$^{002,007}$, 
Richard~W.~Pogge$^{004}$, 
Woong-Tae~Kim$^{012}$
\\
(The KMTNet Collaboration),\\
Fumio~Abe$^{013}$,                   
Richard~Barry$^{014}$,               
David~P.~Bennett$^{014,015}$,       
Aparna~Bhattacharya$^{014,015}$,    
Martin~Donachie$^{016}$,             
Hirosane~Fujii$^{013}$,              
Akihiko~Fukui$^{017,018}$,          
Yoshitaka~Itow$^{013}$,              
Yuki~Hirao$^{019}$,                  
Rintaro~Kirikawa$^{019}$,
Iona~Kondo$^{019}$,                   
Naoki~Koshimoto$^{020,021}$,         
Man~Cheung~Alex~Li$^{016}$,           
Yutaka~Matsubara$^{013}$,             
Yasushi~Muraki$^{013}$,               
Shota~Miyazaki$^{019}$,               
Masayuki~Nagakane$^{019}$,            
Cl\'ement~Ranc$^{014}$,                
Nicholas~J.~Rattenbury$^{016}$,       
Yuki~Satoh$^{019}$,                   
Hikaru~Shoji$^{019}$,                 
Haruno~Suematsu$^{019}$,              
Takahiro~Sumi$^{019}$,                
Daisuke~Suzuki$^{022}$,               
Yuzuru~Tanaka$^{019}$,
Paul~J.~Tristram$^{023}$,             
Tsubasa~Yamawaki$^{019}$,             
Atsunori~Yonehara$^{024}$\\                    
(The MOA Collaboration),\\  
}


\affil{$^{001}$ Department of Physics, Chungbuk National University, Cheongju 28644, Republic of Korea; cheongho@astroph.chungbuk.ac.kr} 
\affil{$^{002}$ Korea Astronomy and Space Science Institute, Daejon 34055, Republic of Korea} 
\affil{$^{003}$ Max Planck Institute for Astronomy, K\"onigstuhl 17, D-69117 Heidelberg, Germany} 
\affil{$^{004}$ Department of Astronomy, Ohio State University, 140 W. 18th Ave., Columbus, OH 43210, USA} 
\affil{$^{005}$ Institute of Natural and Mathematical Sciences, Massey University, Auckland 0745, New Zealand}
\affil{$^{006}$ University of Canterbury, Department of Physics and Astronomy, Private Bag 4800, Christchurch 8020, New Zealand} 
\affil{$^{007}$ Korea University of Science and Technology, 217 Gajeong-ro, Yuseong-gu, Daejeon, 34113, Republic of Korea} 
\affil{$^{008}$ Department of Particle Physics and Astrophysics, Weizmann Institute of Science, Rehovot 76100, Israel}
\affil{$^{009}$ Center for Astrophysics $|$ Harvard \& Smithsonian 60 Garden St., Cambridge, MA 02138, USA} 
\affil{$^{010}$ Department of Astronomy and Tsinghua Centre for Astrophysics, Tsinghua University, Beijing 100084, China} 
\affil{$^{011}$ School of Space Research, Kyung Hee University, Yongin, Kyeonggi 17104, Republic of Korea} 
\affil{$^{012}$ Department of Physics \& Astronomy, Seoul National University, Seoul 08826, Republic of Korea}
\affil{$^{013}$ Institute for Space-Earth Environmental Research, Nagoya University, Nagoya 464-8601, Japan}
\affil{$^{014}$ Code 667, NASA Goddard Space Flight Center, Greenbelt, MD 20771, USA}
\affil{$^{015}$ Department of Astronomy, University of Maryland, College Park, MD 20742, USA}
\affil{$^{016}$ Department of Physics, University of Auckland, Private Bag 92019, Auckland, New Zealand}
\affil{$^{017}$ Instituto de Astrof\'isica de Canarias, V\'ia L\'actea s/n, E-38205 La Laguna, Tenerife, Spain}
\affil{$^{018}$ Department of Earth and Planetary Science, Graduate School of Science, The University of Tokyo, 7-3-1 Hongo, Bunkyo-ku, Tokyo 113-0033, Japan}
\affil{$^{019}$ Department of Earth and Space Science, Graduate School of Science, Osaka University, Toyonaka, Osaka 560-0043, Japan}
\affil{$^{020}$ Department of Astronomy, Graduate School of Science, The University of Tokyo, 7-3-1 Hongo, Bunkyo-ku, Tokyo 113-0033, Japan}
\affil{$^{021}$ National Astronomical Observatory of Japan, 2-21-1 Osawa, Mitaka, Tokyo 181-8588, Japan}
\affil{$^{022}$ Institute of Space and Astronautical Science, Japan Aerospace Exploration Agency, 3-1-1 Yoshinodai, Chuo, Sagamihara, Kanagawa, 252-5210, Japan}
\affil{$^{023}$ University of Canterbury Mt.\ John Observatory, P.O. Box 56, Lake Tekapo 8770, New Zealand}
\affil{$^{024}$ Department of Physics, Faculty of Science, Kyoto Sangyo University, 603-8555 Kyoto, Japan}

\altaffiltext{101}{KMTNet Collaboration.}
\altaffiltext{102}{MOA Collaboration.}

\begin{abstract}
We present the analysis of a very high-magnification ($A\sim 900$) microlensing event 
KMT-2019-BLG-1953.  A single-lens single-source (1L1S) model appears to approximately 
delineate the observed light curve, but the residuals from the model exhibit small but 
obvious deviations in the peak region.  A binary lens (2L1S) model with a mass ratio 
$q\sim 2\times 10^{-3}$ improves the fits by $\Delta\chi^2=181.8$, indicating that the 
lens possesses a planetary companion.  From additional modeling by introducing an extra 
planetary lens component (3L1S model) and an extra source companion (2L2S model), it is 
found that the residuals from the 2L1S model further diminish, but claiming these 
interpretations is difficult due to the weak signals with $\Delta\chi^2=16.0$ and $13.5$ 
for the 3L1S and 2L2L models, respectively.  From a Bayesian analysis, we estimate that 
the host of the planets has a mass of $M_{\rm host}=0.31^{+0.37}_{-0.17}~M_\odot$ and 
that the planetary system is located at a distance of $D_{\rm L}=7.04^{+1.10}_{-1.33}~{\rm kpc}$ 
toward the Galactic center.  
The mass of the securely detected planet is 
$M_{\rm p}=0.64^{+0.76}_{-0.35}~M_{\rm J}$.
The signal of the potential second planet 
could have been confirmed if the peak of the light curve had been more densely observed 
by followup observations, and thus the event illustrates the need for intensive followup 
observations for very high-magnification events even in the current generation of 
high-cadence surveys.  
\end{abstract}


\section{Introduction}\label{sec:one}

Microlensing events with very high magnifications are of scientific importance 
for various reasons. First, the chance for the lens to pass over the surface of 
the source star is high for these events, and this allows one to measure the angular 
Einstein radius $\thetae$, from which the physical parameters of the lens can be 
better constrained \citep{Gould1994b, Nemiroff1994, Witt1994}.  Second, the detection 
probability is very high for planets located in the lensing zone of the host, and 
thus high-magnification events provide an efficient channel to detect microlensing 
planets \citep{Griest1998}.

Another scientific importance of high-magnification events is that they provide a 
channel to detect multiplanetary systems. The basis for this use of microlensing 
lies in the properties of lensing caustics induced by planets. A planet located in 
the vicinity of the Einstein ring induces two sets of caustics, in which one is 
located away from the host of the planet (planetary caustic) and the other is 
located close to the host (central caustic).  See \citet{Han2006} and \citet{Chung2005} 
for the properties of the planetary and central caustics, respectively.  If a lens has 
multiple planets, the individual planets induce central caustics in the common central 
magnification region and affect the magnification pattern of the region.  For very 
high-magnification events, that are produced by the source passage through the central 
magnification region, then, the chance to detect the signatures of the individual planets 
is high \citep{Gaudi1998}.

The usefulness of the high-magnification channel in detecting multiplanetary systems 
has been demonstrated by the fact that three out of four known microlensing multiplanetary 
systems were detected through this channel.  The first multiplanetary system detected 
through this channel is OGLE-2006-BLG-109L, in which two planets with masses of 
$\sim 0.71~M_{\rm J}$ and $\sim 0.27~M_{\rm J}$ are orbiting around a primary star of 
a mass $\sim 0.50~M_\odot$ with projected orbital separations of $\sim 2.3$~au and 
$\sim 4.6$~au \citep{Gaudi2008, Bennett2016}. This system resembles a scaled version 
of our solar system in that the mass ratio, separation ratio, and equilibrium temperatures 
of the planets are similar to those of Jupiter and Saturn of the Solar system.  The 
microlens OGLE-2012-BLG-0026L is the second system, in which the lens consists of two 
planets with masses of $\sim 0.14~M_{\rm J}$ and $\sim 0.86~M_{\rm J}$ and projected 
separations of $\sim 4.0$~au and $\sim 4.8$~au from the host with about a solar mass 
\citep{Han2013, Beaulieu2016}.  The third system is OGLE-2018-BLG-101L, which is 
composed of two planets with masses $\sim 1.8~M_{\rm J}$ and $\sim 2.8~M_{\rm J}$ 
around a host with a mass $\sim 0.18~M_\odot$.  The system is located at a distance 
of $\sim 7.1$~kpc and it is the farthest system among the known multiplanetary systems 
\citep{Han2019}.  Besides these microlensing multiplanetary systems, \citet{Ryu2020} 
pointed out the possibility that the lens of the lensing event OGLE-2018-BLG-0532 might 
have a second planet although there also existed another interpretation of the signal. The 
multiplanetary system OGLE-2014-BLG-1722L \citep{Suzuki2018} was detected from the 
planetary signals produced by the combination of the planetary and central caustics.

In this paper, we present the analysis of a very high magnification lensing event KMT-2019-BLG-1953.
For the presentation of the analysis, we organize the paper as follows.  In Section~\ref{sec:two}, 
we describe the observations of the event and the data used in the analysis. In 
Section~\ref{sec:three}, we present analysis of the data conducted under various interpretations 
of the event. We estimate the angular Einstein radius in Section~\ref{sec:four}  and estimate 
the physical lens parameters in Section~\ref{sec:five}.  In Section~\ref{sec:six}, we discuss 
the importance of followup observations for extreme lensing events for both planet detections 
and physical lens parameter determinations.  In Section~\ref{sec:seven}, we summarize the results 
of the analysis and conclude.

\begin{figure}
\includegraphics[width=\columnwidth]{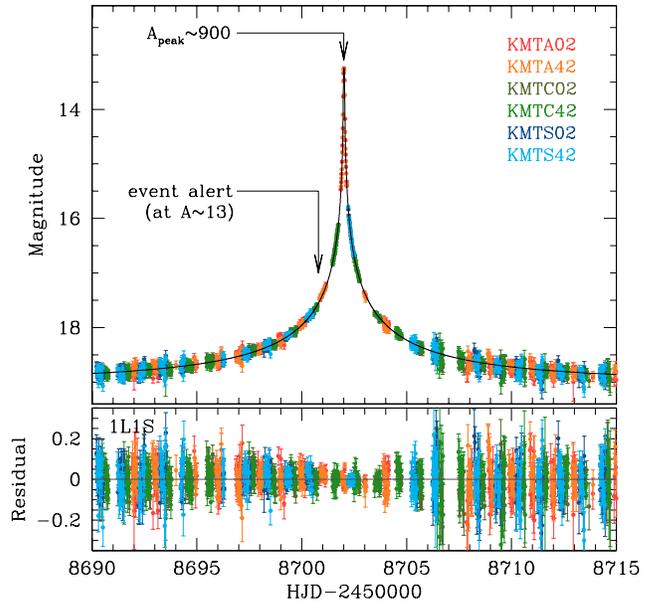}
\caption{
Lightcurve of KMT-2019-BLG-1953.  The curve superposed on the data points is the model 
based on a 1L1S interpretation considering finite-source effects and the lower panel 
shows the residual from the model.  Telescopes used to acquire the data are marked in 
the legend, and the colors of the individual telescopes and data points are chosen to 
match one another.  
\smallskip
}
\label{fig:one}
\end{figure}

\section{Observation and Data}\label{sec:two}

The lensing event KMT-2019-BLG-1953 occurred on a star located toward the
Galactic bulge field. The equatorial coordinates of the lensed star (source) are
$({\rm R.A.}, {\rm decl.})_{\rm J2000}=(17:56:27.90, -28:12:04.00)$.
The corresponding Galactic coordinates are
$(l,b) = (1^\circ\hskip-2pt .85, -1^\circ\hskip-2pt .67)$.

\begin{figure*}
\epsscale{0.75}
\plotone{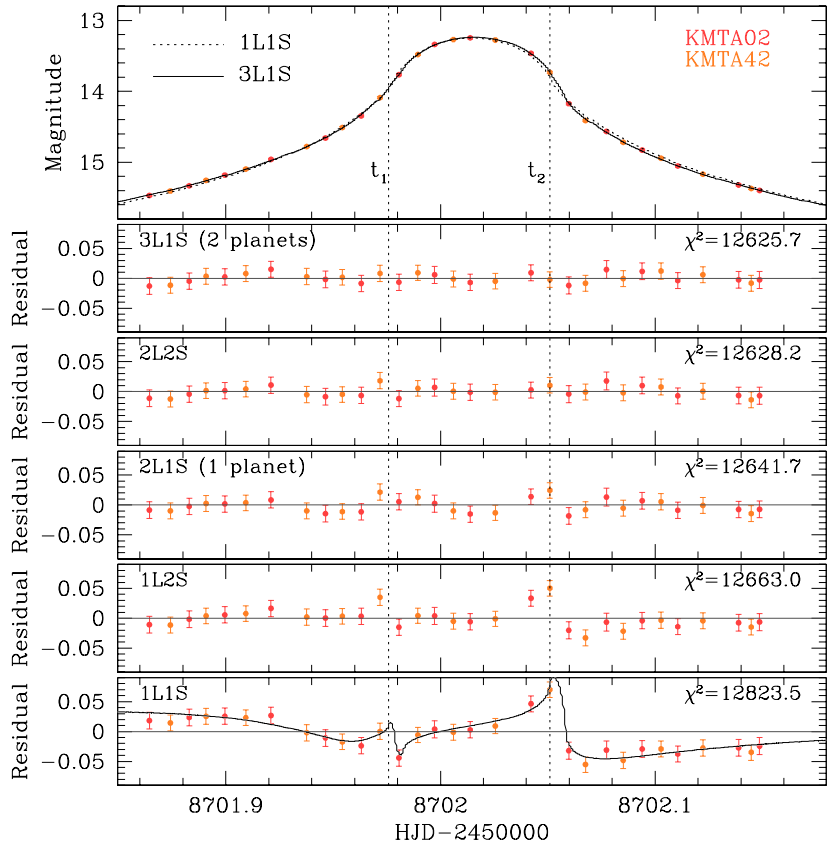}
\caption{
Zoomed-in view around the peak region of the light curve (top panel).  Plotted over 
the data points are the model curves of the 3L1S (solid curve) and 1L1S (dotted curve) 
solutions.  In the five bottom panels, we present the residuals from the five tested 
models based on the 3L1S, 2L2S, 2L1S, 1L2S, and 1L1S interpretations and mark the 
$\chi^2$ value of the fits.  The two times marked by $t_1=8701.975$ and $t_2=8702.051$ 
correspond to the two epochs at which the two caustic-involved bumps in the residuals 
from the 1L1S model arise.  The curve in the bottom panel represents the difference 
between the 3L1S and 1L1S models.
\smallskip
}
\label{fig:two}
\end{figure*}

The magnification of the source flux induced by lensing was first found by the 
Korea Microlensing Telescope Network (KMTNet) survey \citep{Kim2016, Kim2018} 
on 2019-08-05 (${\rm HJD}^\prime\equiv {\rm HJD}-2450000\sim 8701$) when the 
magnification of the source flux was $A\sim 13$. The KMTNet survey was conducted 
utilizing three identical 1.6~m telescopes that were globally located at the Siding 
Spring Observatory in Australia (KMTA), Cerro Tololo Interamerican Observatory in 
Chile (KMTC), and the South African Astronomical Observatory in South Africa (KMTS). 
Each of the KMTNet telescopes was equipped with a camera consisting of four 9k$\times$9k 
chips, yielding 4 deg$^2$ field of view. Images from the survey were mainly taken in the 
$I$ band and a subset of images were obtained in the $V$ band for the source color measurements. 
The event was located in the two overlapping KMTNet fields of BLG02 and BLG42, toward 
which observations were conducted most frequently among the total 27 KMTNet fields.  
Being located in the two overlapping fields in which each field was observed with a 
30~min cadence, the event was observed with a combined cadence of 15 min.  The cadence 
of the $V$-band observations was about one tenth of the $I$-band cadence.

Photometry of the data was conducted using the pipeline developed by \citet{Albrow2009} 
based on the difference imaging method \citep{Tomaney1996, Alard1998}.  For the source 
color measurement, additional photometry was conducted using the pyDIA code 
\citep{Albrow2017} for a subset of the KMTA data set. For the data used in the analysis, 
error-bars from the photometry pipelines were readjusted following the routine described 
in \citet{Yee2012}.

We note that there exist additional data of the event acquired by the Microlensing
Observations in Astrophysics \citep[MOA:][]{Bond2001} survey. The MOA survey found
the event, designated as MOA 2019-BLG-372, two days after the detection by the KMTNet 
survey.  The MOA data are not used in the analysis because (1) the observational cadence 
is low, (2) the peak of the light curve is not covered, and (3) the photometric quality 
of the data is not high.

In Figure~\ref{fig:one}, we present the light curve of the lensing event. It shows that
the source flux is greatly magnified.  From modeling the light curve based on a single-source 
and single-lens (1L1S) interpretation, it is found that the source flux is magnified by 
$A_{\rm peak}\sim 900$ at the peak. We will discuss the modeling in the following section.  
Figure~\ref{fig:two} shows the zoomed-in view around the peak region of the light curve, which 
shows the deviation affected by finite-source effects. The duration of the finite-source 
deviation was about 2 hours.

\begin{deluxetable*}{lcccc}
\tablecaption{Lensing parameters of 1L1S, 2L1S, 1L2S, and 2L2S models\label{table:one}}
\tablewidth{480pt}
\tablehead{
\multicolumn{1}{c}{Parameter}   &
\multicolumn{1}{c}{1L1S}        &
\multicolumn{1}{c}{2L1S}        &
\multicolumn{1}{c}{1L2S}        &
\multicolumn{1}{c}{2L2S}        
}
\startdata                                              
$\chi^2/{\rm dof}$              &  12823.5/12621        &  12641.7/12618         &   12663.0/12617         &  12628.2/12614         \\
$t_0$ (${\rm HJD}^\prime´$)     &  $8702.015\pm 0.001$  &  $8702.016\pm 0.001$   &   $8702.015\pm 0.001$   &  $8702.014\pm 0.001$   \\
$t_{0,2}$ (${\rm HJD}^\prime´$) &  --                   &  --                    &   $8701.949\pm 0.012$   &  $8702.042\pm 0.010$   \\
$u_0$ ($10^{-3}$)               &  $0.70\pm 0.04$       &  $0.72\pm 0.04$        &   $0.02\pm 0.12$        &  $0.71\pm 0.06$        \\
$u_{0,2}$ ($10^{-3}$)           &  --                   &  --                    &   $7.49\pm 1.16$        &  $1.10\pm 0.90$        \\
$t_{\rm E}$ (days)              &  $16.60\pm 0.25$      &  $16.05\pm 0.23$       &   $15.98\pm 0.27$       &  $16.01\pm 0.25$       \\
$s$                             &  --                   &  $2.51\pm 0.31$        &   --                    &  $2.08\pm 0.25$        \\
$q$ ($10^{-3}$)                 &  --                   &  $1.97\pm 0.63$        &   --                    &  $1.37\pm 0.61$        \\
$\alpha$ (rad)                  &  --                   &  $2.408\pm 0.038$      &   --                    &  $2.494\pm 0.028$      \\
$\rho$ ($10^{-3}$)              &  $2.32\pm 0.04$       &  $2.37\pm 0.04$        &   $2.31\pm 0.04$        &  $2.33\pm 0.05$        \\
$\rho_2$ ($10^{-3}$)            &  --                   &  --                    &   $7.87\pm 2.93$        &  $0.65\pm 0.33$        \\
$q_F$                           &  --                   &  --                    &   $0.108\pm 0.009$      &  $0.079\pm 0.024$      
\enddata                            
\tablecomments{
${\rm HJD}^\prime\equiv {\rm HJD}-2450000$.
\smallskip
}
\end{deluxetable*}

\section{Modeling Lightcurve}\label{sec:three}

\subsection{1L1S Modeling}\label{sec:three-one}

Considering the apparently smooth and symmetric shape, we first model the observed light 
curve with a 1L1S interpretation. Modeling is carried out by searching for the lensing 
parameters that best describe the observed light curve. A 1L1S lensing light curve 
affected by finite-source effects is described by four lensing parameters. These 
parameters include $t_0$, $u_0$, $t_{\rm E}$, and $\rho$, which represent the time 
of the closest lens-source approach, the lens-source separation at that time (impact 
parameter), the event timescale, and the normalized source radius, respectively. The 
lensing parameters are searched for using a downhill approach based on the MCMC method. 
In computing finite-source magnifications, we use the semi-analytic expression that was 
derived by \citet{Gould1994a} and later expanded by \citet{Yoo2004} to consider the 
variation of the source star's surface brightness caused by limb darkening. We choose 
the limb-darkening coefficients from the table of \citet{Claret2000} based on the 
source type. The procedure for determining the source type will be discussed in 
Section~\ref{sec:four}.

In Table~\ref{table:one}, we present the best-fit lensing parameters of the 1L1S 
model. To be noted among the lensing parameters is that the impact parameter of the 
lens-source approach, $u_0=(0.70\pm 0.04)\times 10^{-3}$, is extremely small, resulting 
in a very high lensing magnification.  In Figure~\ref{fig:two}, we present the 1L1S model 
curve (dotted curve in the top panel) in the peak region of the light curve. The residuals 
from the model are shown in the bottom panel. The 1L1S model appears to approximately 
delineate the observed light curve, but a close inspection of the residuals reveals that 
the model exhibits small but obvious deviations with $\Delta I\lesssim 0.07$~mag in the 
peak region.  From an additional modeling considering {\it annual} microlens parallax effects 
\citep{Gould1992}, it is found that the microlens parallax $\pie$ cannot be measured, mainly 
due to the relatively short timescale, $t_{\rm E}\sim 16$~days, of the event.  It is known 
that {\it terrestrial} parallax effects can be detected for events with extreme magnifications 
\citep{Gould1997, Gould2009}, and thus we also check the model considering these effects.  
From this, we find that $\pie$ cannot be securely measured mainly because the peak of the 
light curve is covered by only a single observatory, i.e., KMTA.

\subsection{2L1S Modeling}\label{sec:three-two}

Considering that a companion to a lens can induce deviations in the peak region of a very 
high-magnification event, we check whether the deviation from the 1L1S model can be explained 
by the existence of a binary companion to the lens. In order to check this possibility, we 
additionally conduct binary-lens (2L1S) modeling. Adding one more lens component in a lensing 
modeling requires including additional lensing parameters. These parameters are the projected 
separation between the lens components, $s$ (normalized to $\thetae$), the mass ratio between 
the lens components, $q=M_2/M_1$, and the source trajectory angle as measured from the binary 
axis, $\alpha$ (source trajectory angle). In the 2L1S modeling, we divide the lensing parameters 
into two groups.  The grid parameters $s$ and $q$ in the first group are searched for using a 
grid search approach, while the remaining parameters are searched for using a downhill approach 
based on the MCMC method. In the first-round modeling, we construct $\Delta\chi^2$ maps in the 
grid-parameter space and investigate the maps to check the existence of local minima that result 
in possible degenerate solutions. 
In the second-round modeling, we refine the individual local 
minima by allowing $s$ and $q$ parameters to vary and find a global solution by comparing the 
$\chi^2$ values of the local solutions.

\begin{figure}
\includegraphics[width=\columnwidth]{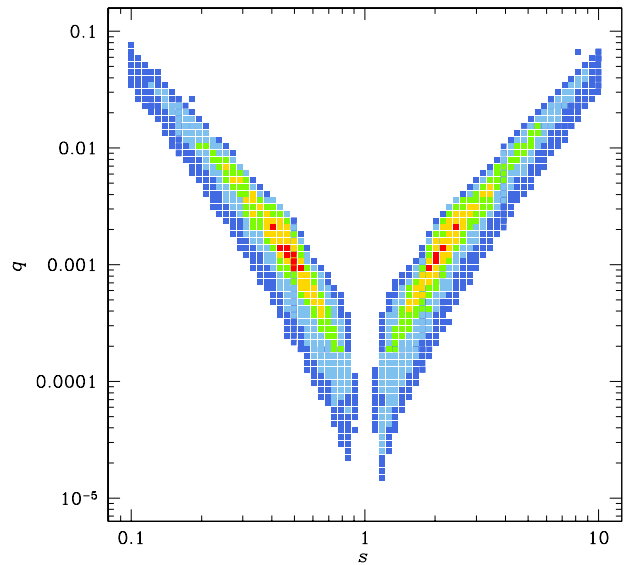}
\caption{
Map of $\Delta\chi^2$ on the $s$--$q$ parameter plane obtained from the grid searches for 
the parameters based on the 2L1S modeling.  Points with different colors denote the regions 
with $<1n\sigma$ (red), $<2n\sigma$ (yellow), $<3n\sigma$ (green), $<4n\sigma$ (cyan), and 
$<5n\sigma$ (blue), where $n=2$.
\smallskip
}
\label{fig:three}
\end{figure}

We find that the 2L1S model substantially reduces the 1L1S residuals.  In Table~\ref{table:one}, 
we present the best-fit lensing parameters of the 2L1S model together with $\chi^2/{\rm dof}$.  
Here ``dof'' denotes the degree of freedom.  The measured mass ratio between the binary lens 
components is $q\sim 2\times 10^{-3}$, indicating that the companion to the lens is a planetary 
mass object. The 2L1S solution is subject to the well-known close/wide degeneracy \citep{Griest1998, 
Dominik1999, An2005}.  The presence of the degenerate solutions are shown in the $\Delta\chi^2$ map 
on the $s$--$q$ plane, shown in Figure~\ref{fig:three}, constructed from the first-round grid search 
for solutions.  The map shows that there are two locals with $s<1.0$ (close solution) and $s>1.0$ 
(wide solution).  The presented parameters in Table~\ref{table:one} are for the solution with $s>1.0$, 
and the solution with $s<1.0$ have similar parameters except $s_{\rm close}\sim s_{\rm wide}^{-1}$.  
In Figure~\ref{fig:two}, we present the residuals from the 2L1S solution with $s>1.0$.  The 2L1S 
model improves the fit by $\Delta\chi^2=181.8$, indicating that the planet is firmly detected.  
From the inspection of the residuals, however, it is found that the 2L1S residuals still exhibit 
subtle deviations from the model.  This hints that the 2L1S solution may not be adequate to fully 
explain the central deviation.

\begin{figure}
\includegraphics[width=\columnwidth]{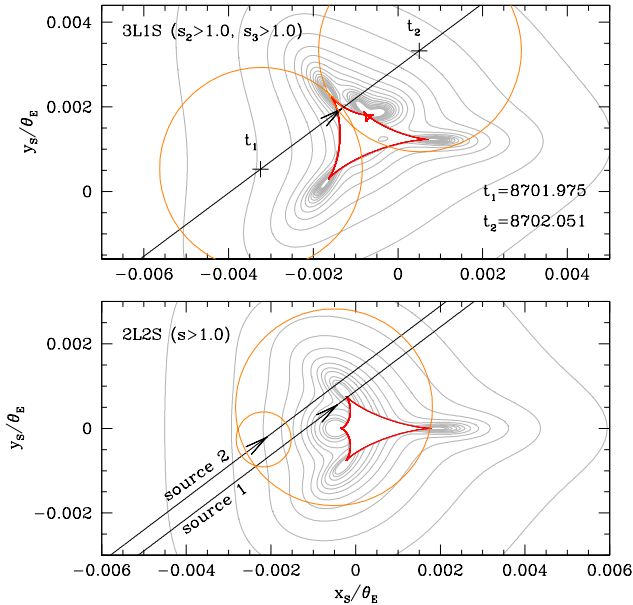}
\caption{
Lens system configurations of the 3L1S (upper panel) and 2L2S (lower panel) models.  In 
each panel, the red cuspy figure represents the caustic, and the line with an arrow denotes 
the source trajectory.  For the 2L2S model, there are two source trajectories corresponding 
to the individual source stars.  The two orange circles in the upper panel represent the 
source positions at the two epochs marked by $t_1$ and $t_2$ in the light curve presented 
in Fig.~\ref{fig:two}. The orange circles in the lower panel represent the positions of 
two source stars at $t_0$. The size of the source circle is scaled to the caustic size. 
Lengths are scaled to the angular Einstein radius corresponding to the total mass of the 
lens.  The gray curves around the caustic represent equi-magnification contours.
\smallskip
}
\label{fig:four}
\end{figure}

\subsection{1L2S and 2L2S Modeling}\label{sec:three-three}

We also check the possibility that the source is a binary (2S). We first test a model in
which the lens is a single object and the source is a binary: 1L2S model. Similar to the
2L1S case, a 1L2S modeling requires extra lensing parameters in addition to those of a
1L1S modeling.  Following the parameterization of \citet{Hwang2013}, these additional 
parameters are $t_{0,2}$, $u_{0,2}$, $\rho_2$, and $q_F$, which represent the time of the 
closest lens approach to the source companion, the lens-companion separation at $t_{0,2}$, 
the normalized radius of the companion source star, and the flux ratio between the two source 
stars, respectively.  In the first-round modeling, we set the initial parameters related to 
the first source ($t_0$, $u_0$, $t_{\rm E}$, and $\rho$) as those determined from the 1L1S 
model and test various trajectories of the second source. In the second round, we refine the 
solutions by letting all parameters vary. The best-fit lensing parameters of the 1L2S solution 
are presented in Table~\ref{table:one} and the residuals from the solution are shown in 
Figure~\ref{fig:two}.  It is found that the 1L2S solution improves the fit by 
$\Delta\chi^2=160.5$ with respect to the 1L1S solution, but the fit is worse than the 2L1S 
solution by $\Delta\chi^2=21.3$.

We additionally check a model in which both the lens and source are binaries: 2L2S model.
Considering that the 2L1S solution substantially improves the fit, we start modeling with
the initial binary-lens parameters, i.e., $(s, q, \alpha)$, as those of the 2L1S solution.
Considering also that the subtle residuals from the 2L1S solution are confined to the
peak region of the light curve, we test various source trajectories passing close to the 
first source.  In Figure~\ref{fig:two}, we present the residuals of the 2L2S solution, 
and we list the lensing parameters in Table~\ref{table:one}. We note that the model is 
subject to the close/wide degeneracy in $s$ and the presented parameters are for the wide 
solution with $s>1.0$.  In the lower panel of Figure~\ref{fig:four}, we present the lens 
system configuration, in which the source trajectories of the two source stars with respect 
to the caustic in the central magnification region is shown.  It is found that the 2L2S model 
further reduces the residuals from the 2L1S model. The improvement of the fit is 
$\Delta\chi^2=13.5$ relative to the 2L1S model.

\begin{deluxetable*}{lcccc}
\tablecaption{Lensing parameters of 3L1S model\label{table:two}}
\tablewidth{480pt}
\tablehead{
\multicolumn{1}{c}{Parameter}            &
\multicolumn{1}{c}{close-close}          &
\multicolumn{1}{c}{close-wide}           &
\multicolumn{1}{c}{wide-close}           &
\multicolumn{1}{c}{wide-wide}            \\  
\multicolumn{1}{c}{}                     &
\multicolumn{1}{c}{($s_2<1.0$, $s_3<1$)}   &
\multicolumn{1}{c}{($s_2<1.0$, $s_3>1$)}   &
\multicolumn{1}{c}{($s_2>1.0$, $s_3<1$)}   &
\multicolumn{1}{c}{($s_2>1.0$, $s_3>1$)}        
}
\startdata                                              
$\chi^2/{\rm dof}$           &  12628.5/12525         &  12626.6/12525         &  12627.9/12525         &  12625.7/12525         \\
$t_0$ (${\rm HJD}^\prime´$)  &  $8702.015\pm 0.002$   &  $8702.014\pm 0.002$   &  $8702.015\pm 0.001$   &  $8702.012\pm 0.001$   \\
$u_0$ ($10^{-3}$)            &  $1.04\pm 0.24$        &  $2.17\pm 0.15$        &  $1.22\pm 0.09$        &  $2.36\pm 0.25$        \\
$t_{\rm E}$ (days)           &  $16.02\pm 0.26$       &  $16.08\pm 0.26$       &  $16.18\pm 0.25$       &  $16.18\pm 0.24$       \\
$s_2$                        &  $0.41\pm 0.10$        &  $0.47\pm 0.07$        &  $2.12\pm 0.13$        &  $2.30\pm 0.36$        \\
$q_2$ ($10^{-3}$)            &  $1.82\pm 1.92$        &  $1.32\pm 1.30$        &  $1.28\pm 0.17$        &  $1.91\pm 0.92$        \\
$\alpha$ (rad)               &  $2.433\pm 0.070$      &  $2.496\pm 0.035$      &  $2.456\pm 0.043$      &  $2.500\pm 0.029$      \\
$s_3$                        &  $0.43\pm 0.12$        &  $4.23\pm 0.63$        &  $0.38\pm 0.05$        &  $4.92\pm 1.14$        \\
$q_3$ ($10^{-3}$)            &  $0.85\pm 1.43$        &  $6.55\pm 1.16$        &  $1.46\pm 0.25$        &  $8.65\pm 2.80$        \\
$\psi$ (rad)                 &  $2.166\pm 0.166$      &  $2.099\pm 0.102$      &  $2.187\pm 0.096$      &  $2.079\pm 0.067$      \\
$\rho$ ($10^{-3}$)           &  $2.41\pm 0.04$        &  $2.40\pm 0.04$        &  $2.40\pm 0.04$        &  $2.40\pm 0.04$        
\enddata                            
\tablecomments{
${\rm HJD}^\prime\equiv {\rm HJD}-2450000$.
\smallskip
}
\end{deluxetable*}

\begin{figure}
\includegraphics[width=\columnwidth]{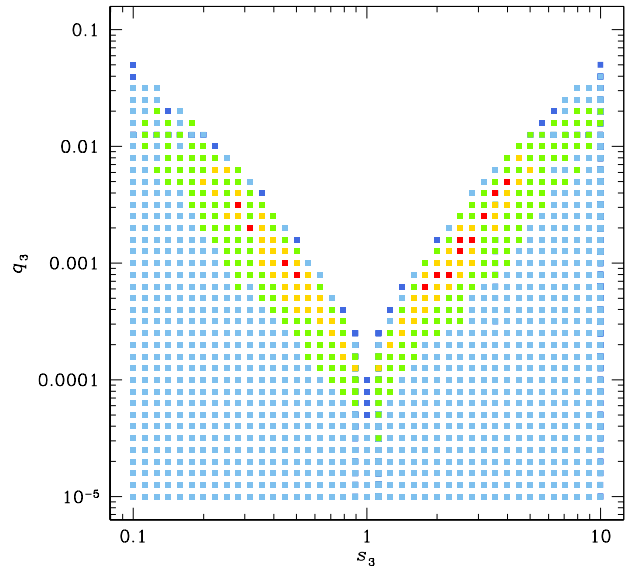}
\caption{
$\Delta\chi^2$ map on the $s_3$--$q_3$ plane obtained from the grid search for these 
parameters based on the 3L1S modeling.  The grid search is conducted with the initial 
values of $(s_2, q_2)$ of the close 2L1S solution.  Color coding of the points is same 
as in Fig.~\ref{fig:three} except that
$n=1$.
\smallskip
}
\label{fig:five}
\end{figure}

\subsection{3L1S Modeling}\label{sec:three-four}

Finally, we test a 3L1S model, in which the lens is composed of two planets and their
host. We test this model because if an additional planet exists, its signal would appear
in the central magnification region, and this may explain the residuals from the 2L1S
model. The addition of a third body, $M_3$, to the binary lens components, $M_1$ and $M_2$,
requires three additional lensing parameters in lens modeling. These parameters are
the projected separation, $s_3$, and mass ratio, $q_3$, between $M_1$ and $M_3$, and the 
orientation angle of $M_3$ with respect to the $M_1$--$M_2$ axis, $\psi$. To designate 
the $M_1$--$M_2$ separation and $M_2/M_1$ mass ratio, we use the notations $s_2$ and $q_2$, 
respectively. The subscript ``1'' is used to designate the host of the planets, and the 
subscripts ``2'' and ``3'' are used to denote the planets. We note that the subscript ``2'' 
is used to designate the planet inducing a larger perturbation in the central magnification 
region. Because a lower-mass planet located close to the Einstein ring of the host can induce 
a larger perturbation than the perturbation induced by a heavier-mass planet located away 
from the Einstein ring, the order of the subscripts ``2'' and ``3'' are not necessarily 
arranged by the mass.

In the 3L1S modeling, we start with the lensing parameters $(s_2, q_2, \alpha)$ of the 
2L1S solution and search for the parameters related to $M_3$, i.e, $(s_3, q_3, \psi)$. 
This strategy is based on the fact that an anomaly induced by two planets, in many cases, 
is dominated by a single planet and the second planet acts as a perturber \citep{Bozza1999, 
Han2001}.  Following this strategy, we first conduct grid searches for $(s_3, q_3, \psi)$ 
parameters by fixing $(s_2, q_2, \alpha)$ parameters and then identify local minima in the 
parameter planes.  Figure~\ref{fig:five} shows the $\Delta\chi^2$ map on the $s_3$--$q_3$ 
plane constructed from this grid searches using the initial values of $(s_2, q_2, \alpha)$ 
of the wide 2L1S solution (with $s_2>1.0$).  In the second round, we refine the individual 
local solutions by allowing all parameters, including $(s_2, q_2, \alpha)$, to vary.

\begin{figure}
\includegraphics[width=\columnwidth]{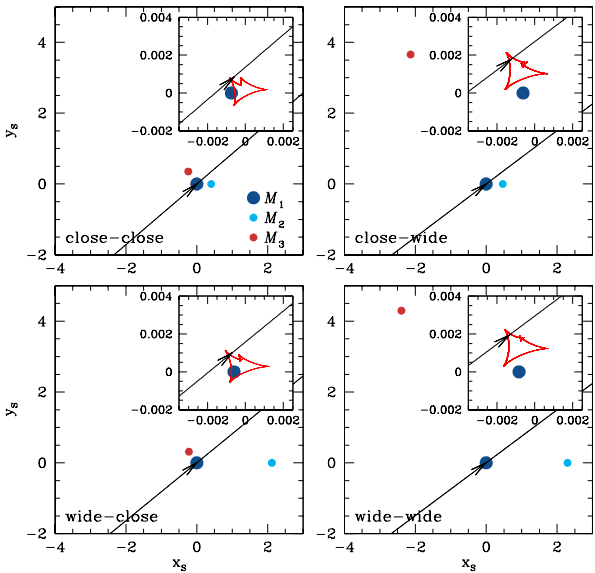}
\caption{
Lens system configurations of the four degenerate 3L1S solutions. The inset in each panel 
shows the zoomed-in view of the central magnification region.
\smallskip
}
\label{fig:six}
\end{figure}

From the 3L1S modeling, we find four sets of degenerate solutions.  The multiplicity of the 
solutions is caused by the close/wide degeneracies in both $s_2$ and $s_3$, and thus the 
individual solutions have $s_2$--$s_3$ pairs of $(s_2<1.0, s_3<1.0)$ (close-close solution), 
$(s_2<1.0, s_3>1.0)$ (close-wide solution), $(s_2>1.0, s_3<1.0)$ (wide-close solution), 
and $(s_2>1.0, s_3>1.0)$ (wide-wide solution), respectively.  
See the $\Delta\chi^2$ map in Figure~\ref{fig:five} showing the two locals 
on the $s_3$--$q_3$ plane with a common values of $(s_2, q_2)$ .
Although the wide-wide solution provides the best fit, the degeneracies among the solutions 
are severe with $\Delta\chi^2\leq 2.8$.  The lensing parameters of the individual solutions 
are presented in Table~\ref{table:two}. The mass ratio of $M_3$ to $M_1$ is in the 
planetary-mass regime regardless of the solutions, with $q_3\sim (0.8-1.5)\times 10^{-3}$ 
and $(6.5-8.7)\times 10^{-3}$ for the solutions with $s_3<1.0$ and $s_3>1.0$, respectively.  
According to the 3L1S solution, then, the lens is a planetary system with two planets.  In 
the four panels of Figure~\ref{fig:six}, we present the lens-system configurations of the 
four degenerate 3L1S solutions. In each panel, the positions of the lens components are 
marked by filled dots and, the inset shows the zoomed-in view of the central caustic.

It is found that the 3L1S model further reduces the residuals from the 2L1S 
solution, improving the fit by $\Delta\chi^2=16.0$ with respect to the 2L1S model.  
In Figure~\ref{fig:two}, we plot the model curve of the 3L1S wide-wide solution (solid 
curve in the top panel) and the residuals from the model. The residuals show that the 
model curve passes through the error bars of all data points around the peak, indicating 
the model well describes the observed light curve.  To show how the 3L1S model explains 
the residuals from the 1L1S model, we draw the curve of the difference between the 3L1S 
and 1L1S solutions in the bottom panel.  For the comparison of the lens system configuration 
with that of the 2L2S solution, we separately present the configuration of the wide-wide 
3L1S solution in the upper panel of Figure~\ref{fig:four}. From the comparison, it is 
found that the right parts of the caustics of the two solutions are similar to each other, 
but the caustic of the 3L1S solution is elongated toward the direction of $M_3$. For the 
3L1S model, the deviations from the 2L1S model at around $t_1$ and $t_2$ are explained by 
the source crossing over the tip of the elongated caustic produced by $M_3$. For the 2L2S 
model, on the other hand, the deviations are explained by the second source's approach 
close to the caustic.

\subsection{Comparison of Models}\label{sec:three-four}

\begin{figure}
\includegraphics[width=\columnwidth]{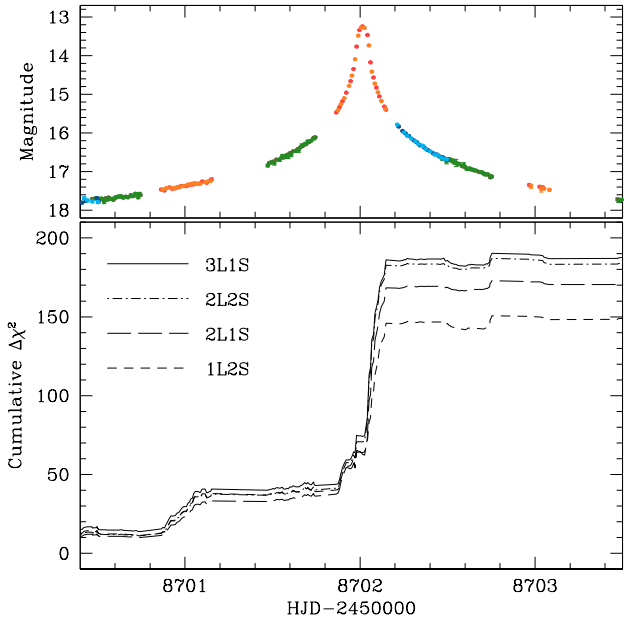}
\caption{
Cumulative distributions of $\Delta\chi^2$ between the tested models (2L1S, 2L2S, and 3L1S 
models) and 1L1S model.  In the upper panel, the observed light curve is presented to show 
the region of fit improvement.
\smallskip
}
\label{fig:seven}
\end{figure}

In Figure~\ref{fig:seven}, we present the cumulative distributions of $\Delta\chi^2$ values 
of the tested models with respect to the 1L1S model.  Results from the comparison of the 
models are summarized as follows.  
\begin{enumerate}
\item
Although the 1L1S solution approximately describes the light curve, the model leaves 
small but obvious deviations in the peak region.  
\item
The 2L1S solution with a planetary-mass companion substantially improves the fit, by 
$\Delta\chi^2=181.8$ with respect to the 1L1S solution.  The 1L2S solution also improves 
the fit, but the model is worse than the 2L1S model by $\Delta\chi^2=21.3$.
\item
With the 3L1S and 2L2S models, the residuals from the 2L1S solution further diminish, and the 
fits improve by $\Delta\chi^2=16.0$ and 13.5 with respect to the 2L1S model, respectively. 
The degeneracy between the 3L1S and 2L2S models is very severe with 
$\Delta\chi^2=\chi_{\rm 2L2S}^2-\chi_{\rm 3L1S}^2 =2.5$,
indicating that it is difficult to distinguish the two models based on only the light curve.
\end{enumerate}

Considering the substantial improvement of the fit by the 2L1S model (single planet model) 
with respect to the 1L1S model, by $\Delta\chi^2=181.8$, the signature of one planet is 
firmly detected.  However, there are two remaining issues in the interpretation of the event.
The first issue is that whether the further improvement of the fit from the 2L1S model with 
the introduction of an extra lens component (second planet), with $\Delta\chi^2=16.0$, 
or an extra source, with $\Delta\chi^2=13.5$, should be seriously considered.  If the signal from 
the extra lens or source is real, then, the second issue is which model, among the 2L2S and 3L1S 
models, is a correct one for the interpretation of the event.

The first issue is closely related to the threshold of detection.  \citet{Dong2009} proposed 
$\Delta\chi^2 \sim 60$ as a threshold (and $\Delta\chi^2 \sim 150$ as a more conservative 
threshold) for the detection of planets through the central perturbations.  Then, although the 
signal of the first planet, with $\Delta\chi^2=181.8$, is firmly detected according to this 
criterion, the signal of the second planet or the source companion does not meet this criterion,
not even considering the increased degrees of freedom.  
Therefore, it is impossible
to claim the 3L1S or 2L2S model for the interpretation of the event considering the unknown 
systematics in the data.

If the extra deviations from the 2L1S model, although weak, is real, we judge that the 3L1S 
model provides a more plausible interpretation of the event than the 2L2S model for two major 
reasons.  First, the signature of the second planet according to the 3L1S solution appears 
in the region where it is expected, i.e., around the peak of a very highly magnified lensing 
event. While this is not really a reason to prefer the 3L1S model, if the opposite were true, 
i.e., the signal from the second planet were coming from somewhere other than the peak, it might 
be a reason to discount the 3L1S model.  The more compelling reason to prefer the 3L1S model is 
that the 2L2S model is physically implausible.  According to the 2L2S model, the projected 
separation (normalized to $\thetae$) between the binary source components during the lensing 
magnification is
\begin{equation}
\Delta u = \left[ 
\left( u_0-u_{0,1}\right)^2 + 
\left( {t_0 - t_{0,2} \over t_{\rm E}}\right)^2
\right]^{1/2}\sim 0.0018.
\label{eq1}
\end{equation}
This corresponds to the physical separation of
\begin{equation}
d_{{\rm S},\perp} = \Delta u D_{\rm S}\thetae \sim 0.0036~{\rm au},
\label{eq2}
\end{equation}
where $D_{\rm S}\sim 8$~kpc denotes the approximate distance to the source and we use $\thetae=0.25$~mas. 
See Section~\ref{sec:four} for the $\thetae$ measurement. The separation is too close for a 
binary system to be stable, and thus the second source would have to be projected a considerable 
distance in front of or behind the first source star in order to avoid merging of the two source 
stars. Even if the source companion is a bit further away, it would give rise to ``ellipsoidal variation'' 
\citep{Han2006b} and ``xallarap''  effects \citep{Rahvar2009}, but such variations are not seen 
in the light curve. For example, if the orbital radius is three times of the projected separation, 
i.e., $a\sim 3d_{{\rm S},\perp}\sim 0.01$~au and assuming  $\sim 1~M_\odot$ of the binary source,
the orbital period would be
\begin{equation}
T=\left[ (a/{\rm au})^3\over M/M_\odot\right]^{1/2}\sim 0.37~{\rm days}.
\label{eq3}
\end{equation}
Then, there would be substantial oscillation in the lensing light curve caused by the ellipsoidal 
variations and xallarap effects. The data quality is good enough to see these variations, if 
existed, during about 10 days around the peak. Therefore, such solutions require extreme 
projection, and thus they are implausible.

\begin{figure}
\includegraphics[width=\columnwidth]{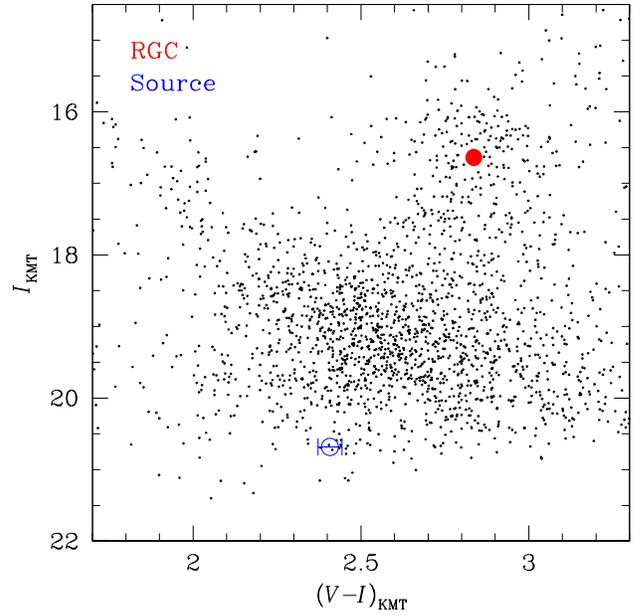}
\caption{
Source location in the instrumental color-magnitude diagram constructed using the pyDIA photometry 
of KMTA $I$- and $V$-band data sets.  The red dot indicates the centroid of the red giant clump (RGC).
\bigskip
}
\label{fig:eight}
\end{figure}

\section{Angular Einstein Radius}\label{sec:four}

We estimate the angular Einstein radius from the combination of the normalized source 
radius $\rho$ and the angular source radius $\theta_*$ by
\begin{equation}
\thetae = {\theta_*\over \rho}.
\label{eq4}
\end{equation}
The value of $\rho$ is measured by modeling the peak part of the light curve that is
affected by finite-source effects.  For the measurement of $\thetae$, then it is required 
to estimate $\theta_*$.

We estimate the angular source radius based on the dereddened color $(V-I)_0$ and magnitude 
$I_0$ using the method of \citet{Yoo2004}. Following the method, we first locate the source 
in the instrumental (uncalibrated) color-magnitude diagram (CMD) and then calibrate the color 
and magnitude using the known values of the red giant clump (RGC) centroid in the CMD as a 
reference. In Figure~\ref{fig:eight}, we present the locations of the source and RGC
centroid in the instrumental CMD constructed using the pyDIA photometry of the KMTA
$I$- and $V$-band data sets. The instrumental color and magnitude of the source are
$(V-I, I)=(2.41\pm 0.04, 20.68\pm 0.01)$. Using the offsets in color and magnitude, $\Delta(V-I, I)$,
from those of the RGC centroid, located at $(V-I, I)_{\rm RGC}=(2.84, 16.64)$, the dereddened
color and magnitude of the source are estimated as
$(V-I, I)_0 = (V-I, I)_{{\rm RGC},0} + \Delta(V-I, I) = (0.63\pm 0.04, 18.39 \pm 0.01)$,
where $(V-I, I)_{{\rm RGC},0} =(1.06, 14.35)$ are the known values of the dereddened color and
magnitude of the RGC centroid 
\citep{Bensby2013, Nataf2013}.
The estimated color and magnitude indicate that the
source is a very late F-type main-sequence star.

\begin{deluxetable}{lc}
\tablecaption{Angular source radius, angular Einstein radius, and relative lens-source proper 
motion\label{table:three}}
\tablewidth{240pt}
\tablehead{
\multicolumn{1}{c}{Quantity}            &
\multicolumn{1}{c}{Value}        
}
\startdata                                              
$\theta_*$ ($\mu$as)   &  $0.61 \pm 0.05$   \\
$\thetae$ (mas)        &  $0.25 \pm 0.02$   \\
$\mu$ (mas~yr$^{-1}$)  &  $5.70 \pm 0.46$   
\enddata                            
\tablecomments{
$\theta_*$: angular source radius, $\thetae$: angular Einstein radius, 
$\mu$: relative lens-source proper motion.
\smallskip
}
\end{deluxetable}

With the measured $(V-I)_0$ and $I_0$, the angular radius of the source is estimated 
first by converting $V-I$ into $V-K$ using the color-color relation of \citet{Bessell1988}
and then using the $(V-K)/\theta_*$ relation of \citet{Kervella2004}. This procedure yields 
the angular source radius of
\begin{equation}
\theta_* = 0.61 \pm 0.05~\mu{\rm as}.
\label{eq5}
\end{equation}
With the normalized source radius, the angular Einstein radius is estimated as
\begin{equation}
\thetae = 0.25 \pm 0.02~{\rm mas}.
\label{eq6}
\end{equation}
Together with the measured event timescale $t_{\rm E}$, the relative lens-source proper 
motion is estimated as
\begin{equation}
\mu = {\thetae \over t_{\rm E}} = 5.70 \pm 0.46~{\rm mas}~{\rm yr}^{-1}.
\label{eq7}
\end{equation}
To be noted is that the measured relative lens-source proper motion is similar to those of 
typical lensing events produced by either bulge or disk lenses that magnify background bulge 
source stars.  In Table~\ref{table:three}, we summarize the values of $\theta_*$, $\thetae$, 
and $\mu$.
We note that the values of $\thetae$ and $\mu$ are estimated based on the parameters of the 
2L1S solution, because the second planet according to the 
3L1S solution is not firmly detected although the solution provides a slightly better fit.

\begin{figure}
\includegraphics[width=\columnwidth]{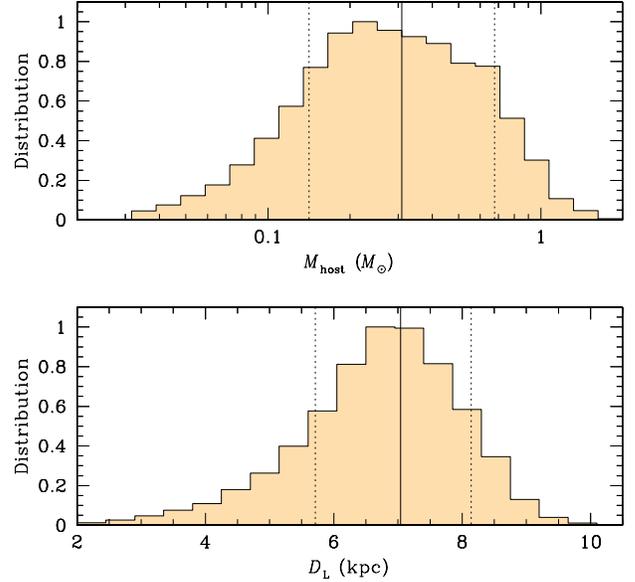}
\caption{
Probability distributions of the mass of the planet host, $M_{\rm host}$ (upper panel)
and the distance to the lens, $D_{\rm L}$ (lower panel). The solid line in each panel 
indicates the median value, and the dotted lines represent the 1$\sigma$ range of the 
distribution.
\smallskip
}
\label{fig:nine}
\end{figure}

\section{Physical Lens Parameters}\label{sec:five}

The mass, $M$, and distance, $D_{\rm L}$, to the lens are uniquely determined by measuring 
the angular Einstein radius $\thetae$ and the microlens parallax, 
$\pie$, i.e.,
\begin{equation}
M={\thetae\over \kappa\pie};\qquad
d_{\rm L}= {{\rm au}\over \pie\thetae+ \pi_{\rm S}}.
\label{eq8}
\end{equation}
Here $\kappa=4G/(c^2{\rm au)}$,
$\pi_{\rm S}={\rm au}/D_{\rm S}$ denotes the parallax of the source, and $D_{\rm S}$ 
indicates the distance to the source.  For KMT-2019-BLG-1953, $\thetae$ is well measured, 
but the short duration of the event makes 
it difficult to measure $\pie$ by the usual method of detecting light curve deviations 
caused by the orbital motion of the Earth: annual microlens parallax \citep{Gould1992}. 
The KMTNet alert (August 5) was issued one week after the final upload (July 29) of the 
``Space-Based Microlens Parallax Survey'' conducted using the {\it Spitzer telescope} 
\citep{Yee2015}, and thus $\pie$ could not be measured through the space-based parallax 
channel \citep{Refsdal1966, Gould1994b}.  Such high magnification events can in principle 
yield terrestrial parallax measurements \citep{Gould1997, Gould2009,Yee2009}, but this 
generally requires that they should be observed near peak from two well-separated observatories. 
However, the peak of KMT-2019-BLG-1953 was only observed from KMTA, and thus $\pie$ could 
not be securely measured through the terrestrial-parallax channel.  Not being able to 
determine $\pie$, we estimate the mass and location of the lens by conducting a Bayesian 
analysis based on the measured $t_{\rm E}$ and $\thetae$ and using the prior models of 
the mass function and the physical and dynamical distributions of lens objects.

For the prior distributions, we adopt the \citet{Han2003} model for the physical
lens distribution and the non-rotating barred bulge model of \citet{Han1995} for
the model of the relative lens-source motion. For the mass function, we adopt the
\citet{Chabrier2003} model for stellar lenses and \citet{Gould2000} model for remnant 
lenses, i.e., white dwarfs, neutron stars, and black holes. With these prior distributions, 
we produce $4\times 10^7$ artificial events by conducting a Monte Carlo simulation. We then 
construct the probability distribution of the physical lens parameters for events with 
$t_{\rm E}$'s and $\thetae$'s located within the ranges of the measured values.

In Figure~\ref{fig:nine}, we present the probability distributions for the mass of 
the planet host, $M_{\rm host}\equiv M_1$ (upper panel), and the distance to the lens, 
$D_{\rm L}$ (lower panel). It is estimated that the host star has a mass of
\begin{equation}
M_{\rm host}=0.31^{+0.37}_{-0.17}~M_\odot,
\label{eq9}
\end{equation}
and is located at a distance of
\begin{equation}
D_{\rm L}=7.04^{+1.10}_{-1.33}~{\rm kpc}.
\label{eq10}
\end{equation}
Therefore, the host of the planets is a low-mass star located either in the bulge or just 
in front of it in the disk.
The mass of the confirmed planet is 
\begin{equation}
M_{\rm p}=0.64^{+0.76}_{-0.35}~M_{\rm J}.
\label{eq11}
\end{equation}
The mass of the second planet, if exists, is in the ranges of 
$0.1 \lesssim M_3/M_{\rm J}\lesssim 1.1$ 
for solutions with $s_3<1.0$ and 
$1.1 \lesssim M_3/M_{\rm J} \lesssim 6.1$ 
for solutions with $s_3>1.0$.

\section{Discussion}\label{sec:six}

Similar to the case of OGLE-2005-BLG-169Lb, for which the planet was discovered because of 
very dense sampling at the peak \citep{Gould2006}, 
the signal of the potential second planet could have been clearly detected if the peak of the 
light curve had been much more densely observed by followup observations,  although interpreting 
the signal might be subject to various types of degeneracy. 
Although KMTNet issued an alert for KMT-2019-BLG-1953 more than 24 
hours before peak, with real-time updates to its web 
page\footnote{http://kmtnet.kasi.re.kr/ulens/kyuha/internal/2019alert/} every three hours, 
no followup observations were taken.  Here we call attention to the potential value of such 
followup observations in the case of this event, and by extension, to other similar events.

Despite the fact that KMT-2019-BLG-1953 lies in one of KMTNet's three highest-cadence fields, 
with a cadence of 15~min, the coverage was not dense enough to securely detect the second 
planet.  Thus, even though the KMTNet observing strategy was originally designed to capture 
the shortest anomalies, due to Earth-mass planets, it is still not frequent enough to fully 
exploit the very rare extreme magnification events such as KMT-2019-BLG-1953. In the era prior 
to the advent of KMTNet, such high magnification and extremely high magnification events were 
a major channel of planet detection, and they were observed at a much higher cadence 
\citep{Gould2010}.

Indeed, high-cadence observations from multiple well-separated sites led to terrestrial 
parallax measurements for two events, OGLE-2007-BLG-224 \citep{Gould2009} and OGLE-2008-BLG-279 
\citep{Yee2009}.  In the case of KMT-2019-BLG-1953, it is far from clear that such well-separated 
observations of the peak would have yielded a successful terrestrial-parallax measurement. For 
example, if $M\sim 0.3\,M_\odot$ and $D_S - D_L\sim 1\,$kpc, as in our best Bayesian estimate, 
then $\pi_{\rm E}\sim 0.085$ with a resulting projected velocity 
$\tilde v \equiv {\rm au}/\pi_{\rm E} t_{\rm E} \rightarrow 1300\,{\rm km\,s^{-1}}$.  Hence, 
the peaks as observed by two telescopes separated by 2500 km would have been displaced by at 
most 2 seconds. This would be too short to measure reliably. Nevertheless, without a parallax 
measurement, we do not know with certainty that the lens was not much closer, in which case 
it could have been measured.

The main point is that extreme microlensing events such as KMT-2019-BLG-1953 are a rich source 
of information, both about planets (multiplanetary systems) and microlens parallaxes. They occur 
only a few times per season, and they should be followed up with intensive observations, when 
possible, even in the current high-cadence surveys.

\section{Summary and Conclusion}\label{sec:seven}

We analyzed a very high-magnification event KMT-2019-BLG-1953.  The model based on the 1L1S 
interpretation with finite-source effects appeared to approximately describe the observed light 
curve, but the residuals from the model exhibited small but obvious deviations with 
$\Delta I\lesssim 0.07$~mag in the peak region.  A 2L1S model revealed the existence of a 
planetary companion to the lens ($q\sim 2\times 10^{-3}$) with a significant confidence level.  
It was found that additional modeling by introducing a second planetary lens companion or an 
extra source companion further reduced the residuals from the 2L1S model.  However, this extra 
signal is not strong enough to confirm the additional lens or source companion.  From the 
Bayesian analysis conducted based on the measured $t_{\rm E}$ and $\thetae$, it was estimated 
that the host of the planet was a low-mass star with a mass of 
$M_{\rm host}=0.31^{+0.37}_{-0.17}~M_\odot$ and the planetary system was located at a distance 
of $D_{\rm L}=7.04^{+1.10}_{-1.33}~{\rm kpc}$ toward the Galactic center.  The mass of the 
confirmed planet was  in the range of 
$M_{\rm p}=0.64^{+0.76}_{-0.35}~M_{\rm J}$.

\acknowledgments
Work by CH was supported by the grants  of National Research Foundation of Korea 
(2017R1A4A1015178 and 2019R1A2C2085965).
Work by AG was supported by US NSF grant AST-1516842 and by JPL grant 1500811.
AG received support from the European Research Council under the European Union's 
Seventh Framework Programme (FP 7) ERC Grant Agreement n.~[32103].
This research has made use of the KMTNet system operated by the Korea
Astronomy and Space Science Institute (KASI) and the data were obtained at
three host sites of CTIO in Chile, SAAO in South Africa, and SSO in
Australia.
The MOA project is supported by JSPS KAKENHI Grant Number JSPS24253004,
JSPS26247023, JSPS23340064, JSPS15H00781, JP17H02871, and JP16H06287.
YM acknowledges the support by the grant JP14002006.
DPB, AB, and CR were supported by NASA through grant NASA-80NSSC18K0274. 
The work by CR was supported by an appointment to the NASA Postdoctoral Program at the Goddard 
Space Flight Center, administered by USRA through a contract with NASA. NJR is a Royal Society 
of New Zealand Rutherford Discovery Fellow.

\end{document}